\documentclass[aps,prb,twocolumn,showpacs]{revtex4}
\usepackage{graphicx}

\begin{document}

\title{(Sr$_3$Sc$_2$O$_5$)Fe$_2$As$_2$ as a possible parent compound for FeAs-based superconductors}

\author{Xiyu Zhu, Fei Han, Gang Mu, Bin Zeng, Peng Cheng,  Bing Shen,  and Hai-Hu Wen}\email{hhwen@aphy.iphy.ac.cn }

\affiliation{National Laboratory for Superconductivity, Institute of
Physics and Beijing National Laboratory for Condensed Matter
Physics, Chinese Academy of Sciences, P. O. Box 603, Beijing 100190,
China}

\begin{abstract}
A new compound with the FeAs-layers, namely
(Sr$_3$Sc$_2$O$_5$)Fe$_2$As$_2$ (abbreviated as FeAs-32522), was
successfully fabricated. It has a layered structure with the space
group of \emph{I4/mmm}, and with the lattice constants a = 4.069
$\AA$ and c =  26.876 $\AA$. The in-plane Fe ions construct a square
lattice which is close to that of other FeAs-based superconductors,
such as REFeAsO (RE = rare earth elements) and (Ba,Sr)Fe$_2$As$_2$.
However the inter FeAs-layer spacing in the new compound is greatly
enlarged. The temperature dependence of resistivity exhibits a weak
upturn in the low temperature region, but a metallic behavior was
observed above about 60 K. The magnetic susceptibility shows also a
non-monotonic behavior. Interestingly, the well-known resistivity
anomaly which was discovered in all other parent compounds, such as
REFeAsO, (Ba,Sr)Fe$_2$As$_2$ and (Sr,Ca,Eu)FeAsF and associated with
the Spin-Density-Wave (SDW)/structural transition has not been found
in the new system either on the resistivity data or the
magnetization data. This could be induced by the large spacing
distance between the FeAs-planes, therefore the antiferromagnetic
correlation between the moments of Fe ions in neighboring
FeAs-layers cannot be established. Alternatively it can also be
attributed to the self-doping effect between Fe and Sc ions. The
Hall coefficient $R_H$ is negative but strongly temperature
dependent in wide temperature region, which indicates the dominance
of electrical conduction by electron-like charge carriers and
probably a multi-band effect or a spin related scattering effect. It
is found that the magnetoresistance cannot be described by the
Kohler's rule, which gives further support to above arguments.

\end{abstract} \pacs{74.70.Dd, 74.25.Fy, 75.30.Fv, 74.10.+v}
\maketitle

\section{Introduction}
Superconductivity in the FeAs-based systems has received tremendous
attention in last several months with the hope that the
superconducting transition temperature could be raised to a higher
value.\cite{Kamihara2008} Experimentally, it has been found that the
highest $T_c$ is about 55-57 K in the fluorine doped REFeAsO
compound\cite{RenZA55K,WangC} or RE doped (Ca,Sr)FeAsF (RE = rare
earth elements).\cite{CP} Meanwhile the family of the FeAs-based
superconductors has been expanded rapidly. In the system of
(Ba,Sr)$_{1-x}$K$_x$Fe$_2$As$_2$ (denoted as FeAs-122), the maximum
T$_c$ at about 38 K was discovered\cite{BaKparent,Rotter,CWCh}. In
the Li$_x$FeAs (denoted as FeAs-111) system, superconductivity at
about 18 K was found.\cite{LiFeAs,LiFeAsChu,LiFeAsUK} In the
material FeSe$_{1-x}$ which has no toxic arsenic and a more simple
structure\cite{WuMK}, it was shown that the sample became
superconductive at about 8 K. By doping or using a high pressure,
the T$_c$ can be pushed higher. Superconductivity at about 2.2 K has
also been found in a NiP-based layered structure
La$_3$Ni$_4$O$_2$P$_4$ with the inter growth of NiP-1111 and 122
building blocks.\cite{Cava} Very recently, another kind of parent
compounds (Sr,Ca,Eu)FeAsF were
reported,\cite{Han,Tegel,HosonoSrFeAsF,ZhuXYNatMat} which may lead
to high temperature superconductors by doping charges into the
system.

Empirically it is found that the superconducting transition
temperature rises when the inter FeAs-layer spacing distance
$d_{FeAs}$ is increased (F-SmFeAsO: T$_c$=55 K, $d_{FeAs}=8.7 \AA$;
(Sr,Ba)$_{1-x}$K$_x$Fe$_2$As$_2$: T$_c$=38 K, $d_{FeAs}=6.5 \AA$;
Li$_x$FeAs: T$_c$=18 K, $d_{FeAs}=6.4 \AA$) in the FeAs-based
superconductors. Surprisingly this seems to be quite similar to the
case in the cuprates. For example, by using the effective layers of
CuO planes in the cuprates, superconductors with different
structures were discovered. They can be categorized into the
so-called single-layer systems La$_{1-x}$Sr$_x$CuO$_4$ and
Bi$_2$Sr$_2$CuO$_6$ with the maximum T$_c$ of about 38 K, or the
double-layer systems YBa$_2$Cu$_3$O$_7$ and
Bi$_2$Sr$_2$CaCu$_2$O$_8$ with the optimal T$_c$ of about 85-92 K,
or the triple-layer system Bi$_2$Sr$_2$Ca$_2$Cu$_3$O$_{10}$ with the
maximum T$_c$ of about 123 K (at ambient pressure).\cite{Cuprate} As
inspired by the experience in the cuprate superconductors, a larger
separation between the superconducting planes (here the FeAs layers)
may lead to a higher $T_c$. Thus it becomes very interesting to see
whether it is possible to fabricate a compound with larger spacing
distance between the FeAs-planes. In this paper, we report the
discovery of another kind of FeAs-based layered compound
(Sr$_3$Sc$_2$O$_5$)Fe$_2$As$_2$ (abbreviated as FeAs-32522), which
has an inter FeAs-layer spacing distance of about 13.438$\AA$, being
much larger than those in all other parent compounds discovered so
far, such as REFeAsO, FeAs-122, and (Sr,Ca,Eu)FeAsF. The temperature
dependence of resistivity, magnetization, Hall effect and
magneto-resistance all exhibit quite similar behaviors as other
parent compounds, but with the exception that the resistivity
anomaly which was associated with the SDW/structural
transition\cite{Dai} has not been found in present system up to 400
K. Superconductivity may be induced by doping electrons or holes in
this new compound.

\section{Experiment}

The compound Sr$_3$Sc$_2$O$_5$Fe$_2$As$_2$ was found while looking
for iron arsenic analogs of materials with this structure type seen
in related chemical systems.\cite{Kishio} The polycrystalline
samples were fabricated by using a two-step solid state reaction
method\cite{XYZ}. Firstly, SrAs and ScAs powders were obtained by
the chemical reaction method with Sr pieces, Sc pieces and As
grains. Then they were mixed with Sc$_2$O$_3$ (purity 99.9\%), SrO
(purity 99\%), Fe$_2$O$_3$(purity 99.9\%), and Fe powder (purity
99.9\%) in the formula (Sr$_3$Sc$_2$O$_5$)Fe$_2$As$_2$, grounded and
pressed into a pellet shape. The weighing, mixing and pressing
processes were performed in a glove box with a protective argon
atmosphere (the H$_2$O and O$_2$ contents are both below 0.1 PPM).
The pellets were sealed in a quartz tube with 0.2 bar of Ar gas and
followed by a heat treatment at 1000 $^o$C for 40 hours. Then it was
cooled down slowly to room temperature.

The X-ray diffraction (XRD) pattern of our samples was carried out
by a $Mac$-$Science$ MXP18A-HF equipment with $\theta - 2\theta$
scan. The XRD data taken using powder sample was analyzed by the
Rietveld fitting method using the GSAS suite\cite{GsAs}. The
starting parameters for the fitting were taken from isostructural
(Sr$_3$Sc$_2$O$_5$)Cu$_2$S$_2$ and the program will finally find the
best fitting parameters.\cite{Kishio} The dc magnetization
measurements were done with a superconducting quantum interference
device (Quantum Design, SQUID, MPMS7). For the magnetotransport
measurements, the sample was shaped into a bar with the length of 3
mm, width of 2.5 mm and thickness of about 0.8 mm. The resistance
and Hall effect data were collected using a six-probe technique on
the Quantum Design instrument physical property measurement system
(PPMS) with magnetic fields up to 9 T. The electric contacts were
made using silver paste with the contacting resistance below 0.06
$\Omega$ at room temperature. The data acquisition was done using a
dc mode of the PPMS, which measures the voltage under an alternative
dc current and the sample resistivity is obtained by averaging these
signals at each temperature. In this way the contacting thermal
power is naturally removed. The temperature stabilization was better
than 0.1\% and the resolution of the voltmeter was better than 10
nV.

\begin{figure}
\includegraphics[width=9cm]{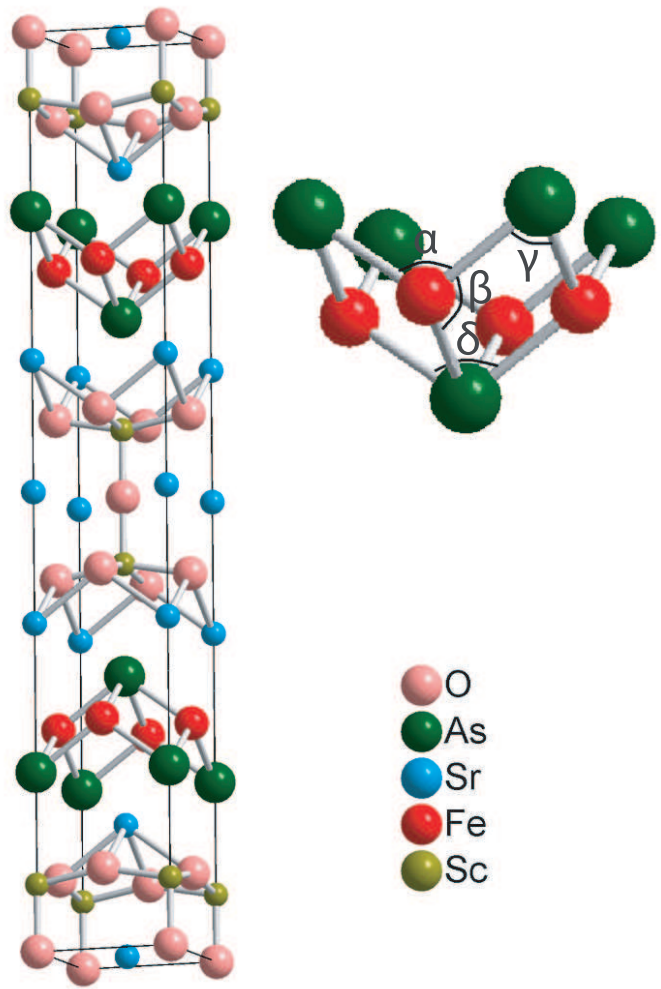}
\caption{(Color online) (Left) Crystal structure of
(Sr$_3$Sc$_2$O$_5$)Fe$_2$As$_2$. Now the Fe$_2$As$_2$ layers are
well separated by the building block Sr$_3$Sc$_2$O$_5$. (Right) An
enlarged view about the FeAs block on which four different angles
are marked: $\alpha$: As-Fe-As (I), $\beta$: As-Fe-As (II),
$\gamma$: Fe-As-Fe (I) and $\delta$: Fe-As-Fe (II).} \label{fig1}
\end{figure}

\begin{figure}
\includegraphics[width=8.5cm]{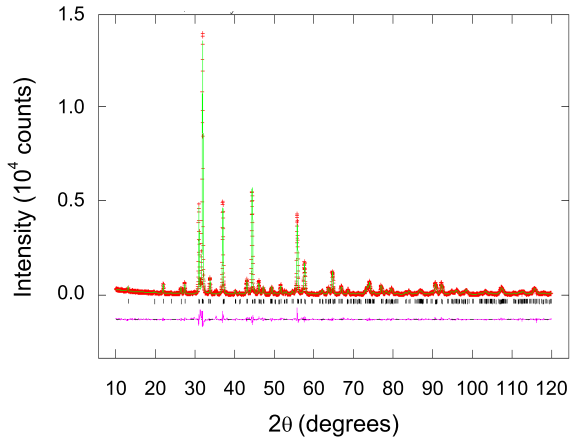}
\caption{(Color online) X-ray diffraction patterns and the Rietveld
fit for the (Sr$_3$Sc$_2$O$_5$)Fe$_2$As$_2$ powder sample.}
\label{fig2}
\end{figure}

\section{Analysis to the structural data}

The XRD pattern for the sample (Sr$_3$Sc$_2$O$_5$)Fe$_2$As$_2$ is
shown in Fig. 2. One can see that the sample is very pure since no
any visible peaks can be detected from the impurity phase. The data
was well fitted with a single tetragonal
(Sr$_3$Sc$_2$O$_5$)Fe$_2$As$_2$ phase with the space group of
\emph{I4/mmm}. Rietveld refinement shown by the solid line in the
figure gives good agreement between the data and the calculated
profiles. Lattice parameters for the tetragonal unit cell was
determined to be a = 4.069 $\AA$ and c = 26.876 $\AA$. In Table I,
the structure parameters, angles of As-Fe-As and Fe-As-Fe (as marked
in Fig.1) were listed with agreement factors: wR$_{p}$ = 11.75$\%$,
R$_{p}$= 8.09$\%$.

\begin{table}
\caption{Fitting parameters, some angles ($^o$) for
(Sr$_3$Sc$_2$O$_5$)Fe$_2$As$_2$. wR$_{p}$ = 11.75$\%$, R$_{p}$=
8.09$\%$.}
\begin{tabular}{cccccccc}
\hline \hline
Atom & site & x & y & z  &B& angle ($^o$)&\\
\hline
 Sc  & 4e  & 0        &     0    &     0.07399(12)&0.28(6) &\\
 Fe  & 4d  &0.5       &     0    &  0.25& 0.46(6)&\\
 Sr  & 2b  &0.5       &     0.5  &     0& 1.2(5)&\\
 Sr  & 4e  &0.5       &     0.5  & 0.14006(6)& 0.51(4)&\\
 As  & 4e  & 0        &     0    &     0.20018(6)& 0.62(4)&\\
 O   & 8g  & 0.5      &     0    &    0.08323(23)&0.54(14)\\
 O   & 2a  & 0        &     0    &     0&0.82(31) &\\
$\alpha$&&&&&& $113.30(7)^o$ \\
$\beta $&&&&&& $107.590(32)^o$\\
$\gamma$&&&&&& $72.410(32)^o$\\
$\delta$&&&&&& $113.30(7)^o$\\

 \hline \hline

\end{tabular}
\label{tab:table1}
\end{table}

It is clear that the a-axis lattice constant of this parent phase is
slightly larger than that of the REFeAsO, (Ba,Sr)Fe$_2$As$_2$ and
(Sr,Ca,Eu)FeAsF systems, while the c-axis one is much
larger.\cite{Kamihara2008,XYZ,Canfield,Han} The skeleton shown in
the right hand side of Fig.1 gives the structure model of our
sample. It is clear that the Fe ions construct a square lattice and
the FeAs-layer stacks with the (Sr$_3$Sc$_2$O$_5$) building block
along c-axis. The spacing distance between two neighboring
FeAs-layers of about 13.438 $\AA$ is much larger than those in all
other FeAs-based compounds discovered so far, such as REFeAsO,
(Ba,Sr)Fe$_2$As$_2$, and (Sr,Ca,Eu)FeAsF. This may induce high $T_c$
superconductivity by doping charges into the system. In fact, this
kind of structures were reported and classified to the general
formula (Cu$_2$S$_2$)(Sr$_{n+1}$M$_n$O$_{3n-1}$).\cite{Kishio}

\begin{figure}
\includegraphics[width=8.5cm]{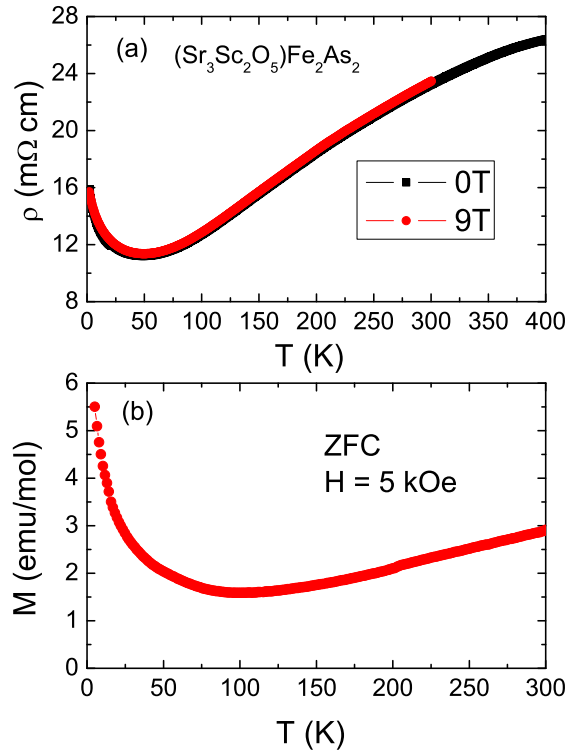}
\caption{(Color online) (a) Temperature dependence of resistivity
for the (Sr$_3$Sc$_2$O$_5$)Fe$_2$As$_2$ sample at zero field up to
400 K and 9 T up to 300 K. A clear upturn in the low temperature
region can be seen. (b) Temperature dependence of dc magnetization
for the zero field cooling (ZFC) process at a magnetic field of $H$
= 5000 Oe. } \label{fig3}
\end{figure}

\section{Electrical and magnetic properties}

In Fig.3 (a) we present the temperature dependence of resistivity
for the (Sr$_3$Sc$_2$O$_5$)Fe$_2$As$_2$ sample under two magnetic
fields 0 T and 9 T. The data under zero field were collected in
temperature region up to 400 K. A weak upturn in the low-temperature
regime can be seen under both fields, representing a weak
semiconductor like behavior for the present sample. This behavior
was attributed to the weak localization effect in our sample.
Interestingly, the similar weak upturn of resistivity was observed
in stoichiometric LaOFeP.\cite{CavaLaOFeP} A metallic behavior can
be seen in the temperature dependence of resistivity in the
temperature regime above 60 K in our sample. Surprisingly, no
resistivity anomaly corresponding to the SDW transition was observed
up to 400 K. This can be further confirmed by the magnetization data
as shown in Fig.3 (b), which shows the zero field cooled dc
magnetization at 5000 Oe. Again no anomaly was observed in the
magnetization data. This is rather different from other FeAs-based
parent compounds, such as REFeAsO\cite{Kamihara2008},
(Ba,Sr)Fe$_2$As$_2$\cite{BaKparent,Rotter} and
(Sr,Ca,Eu)FeAsF\cite{Han}, where clear resistivity and magnetization
anomaly associated with the SDW/structral transition have been
observed. We attribute it to the rather large spacing distance
between the neighboring FeAS-layers in the present system, which
prevents from forming the anti-ferromagnetic correlation between the
moments of Fe ions in neighboring FeAs-layers.

\begin{figure}
\includegraphics[width=9cm]{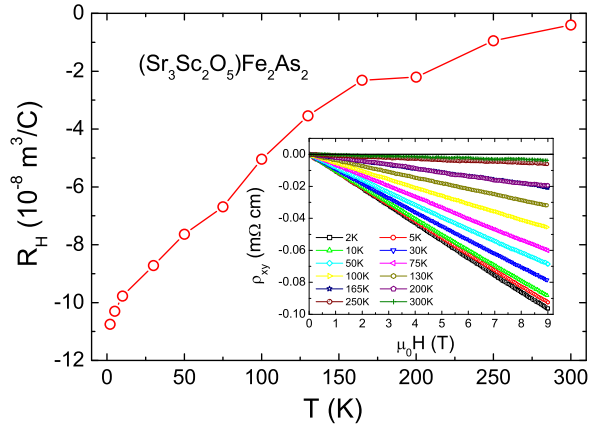}
\caption{(Color online)  Temperature dependence of Hall coefficient
$R_H$ determined on the sample (Sr$_3$Sc$_2$O$_5$)Fe$_2$As$_2$.
Negative values of $R_H$ and a strong temperature dependence can be
seen. Inset: The raw data of the Hall resistivity $\rho_{xy}$ versus
the magnetic field $\mu_0 H$ at different temperatures. }
\label{fig4}
\end{figure}

\begin{figure}
\includegraphics[width=9cm]{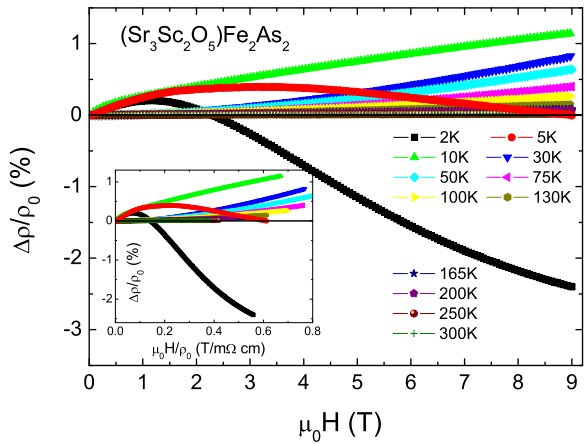}
\caption{(Color online) Field dependence of MR for the present
sample at different temperatures is shown in the main frame. The
inset shows the Kohler plot of MR. The MR is positive at a low field
and turns to be negative at low temperatures. } \label{fig5}
\end{figure}

To further understand the conducting carriers in the present sample,
we also carried out the Hall effect measurements on the present
sample. The inset of Fig. 4 shows the magnetic field dependence of
Hall resistivity ($\rho_{xy}$) at different temperatures. In the
experiment, $\rho_{xy}$ was taken as $\rho_{xy}$ = [$\rho$(+H) -
$\rho$(-H)]/2 at each point to eliminate the effect of the
misaligned Hall electrodes. It is clear that $\rho_{xy}$ is negative
at all temperatures below 300 K for (Sr$_3$Sc$_2$O$_5$)Fe$_2$As$_2$
leading to a negative Hall coefficient $R_H$ =$\rho_{xy}$/H. This is
similar to that of REFeAsO\cite{ChengPPRB,MandrusPRB} and
(Ba,Sr)Fe$_2$As$_2$ parent phases, but in sharp contrast with
(Sr,Ca,Eu)FeAsF in which the positive Hall coefficient R$_H$ was
found. The temperature dependence of the Hall coefficient $R_H$ is
presented in the main frame of Fig. 4. One can see that $R_H$
remains negative in wide temperature regime up to 300 K and the
absolute value of $R_H$ decreases monotonically from 2 K to 300 K.
This indicates that electron-type charge carriers dominate the
conduction in the present sample. The strong temperature dependent
behavior of the Hall coefficient $R_H$ suggests either a strong
multi-band effect or a spin related scattering effect. The absolute
value of $R_H$ is remarkably smaller than that of SrFeAsF system,
indicating a relatively higher density of charge carriers.

The magnetoresistance (MR) is a very powerful tool to investigate
the properties of electronic scattering.\cite{LiQ,YangHall} Field
dependence of MR for the present sample at different temperatures is
shown in the main frame of Fig 5. One can see a systematic evolution
of the curvature in the $\Delta \rho/\rho_0$ vs H curve, where
$\Delta \rho= \rho(H)-\rho_0$, $\rho(H)$ and $\rho_0$ represents the
longitudinal resistivity at a magnetic field $H$ and that at zero
field. The MR data at 2 K is positive with a hump shape below 2.3 T
and becomes negative at higher fields. While at 5 K, a maximum value
of MR appears at about 3.3 T and it decreases until becoming zero at
9 T. The data at 10 K have a downward curvature compared with that
measured at higher temperatures. We attribute the complicated
behavior at low temperatures, especially the negative MR effect at 2
K, to the Kondo effect which is associated very well with the upturn
of resistivity or the magnetic susceptibility, or to the magnetic
field induced delocalization effect. The semiclassical transport
theory has predicted that the Kohler's rule will be held if only one
isotropic relaxation time is present in a solid state
system.\cite{Kohler} The Kohler's rule can be written as

\begin{equation}
\frac{\Delta \rho}{\rho_0}=\frac{\rho(H)-\rho_0}{\rho_0}=F(\frac{
H}{\rho_0}).\label{eq:1}
\end{equation}

This equation means that the $\Delta \rho/\rho_0$ vs $ H/\rho_0$
curves for different temperatures, the so-called Kohler's plot,
should be scaled to a universal curve if the Kohler's rule is
obeyed. The scaling based on the Kohler plot of our sample is
revealed in the inset of Fig.5. An obvious violation of the Kohler's
rule can be seen from this plot. This behavior may indicate a
multi-band effect or a complicated scattering mechanism, like the
Kondo effect or the weak localization in the low temperature region.
This kind of upturn of resistivity and the magnetic susceptibility
in the low temperature limit seem to be common features for the
parent phases of the FeAs-based compounds and further studies are
certainly required to unravel the underlying physics.

\section{Concluding remarks}

In summary, a parent phase of the FeAs-based family,
(Sr$_3$Sc$_2$O$_5$)Fe$_2$As$_2$, with the space group of
\emph{I4/mmm} was synthesized successfully using a two-step solid
state reaction method. Rather large spacing distance between the
neighboring FeAs-layers was found. No anomaly associated with the AF
order was observed either in the data of temperature dependence of
resistivity or dc magnetization, which was attributed to the rather
large spacing distance between the neighboring FeAs-layers. Strong
Hall effect was observed in the temperature region up to 300 K. We
found that the Hall coefficient $R_H$ is negative in the entire
temperature regime. We also observed a delocalization effect at low
temperatures from the MR data. The violation of the Kohler's rule
along with the strong temperature dependence of $R_H$ may suggest a
multi-band and/or a spin scattering effect in this system. By doping
electrons or holes in this new compound, superconductivity may be
achieved.

This work is supported by the Natural Science Foundation of China,
the Ministry of Science and Technology of China (973 project:
2006CB01000, 2006CB921802), the Knowledge Innovation Project of
Chinese Academy of Sciences (ITSNEM).

\end{document}